\documentclass[aps,eqsecnum,twocolumn,amsfonts]{revtex4}
\usepackage{amsfonts}
\usepackage{amsmath}
\usepackage{bm}
\usepackage{epsf}

\newcommand{\beq}{\begin{equation}}
\newcommand{\eeq}{\end{equation}}
\newcommand{\beqa}{\begin{eqnarray}}
\newcommand{\eeqa}{\end{eqnarray}}

\newcommand{\ket}[1]{| #1    \rangle }
\newcommand{\bra}[1]{ \langle   #1  | }
\newcommand{\ave}[1]{  \langle #1   \rangle }
\newcommand{\qave}[2]{  \langle #1  | #2  | #1  \rangle }
\newcommand{\mel}[3]{  \langle #1  | #2   | #3  \rangle }
\newcommand{\amp }[2]{ \langle #1 |  #2  \rangle }
\newcommand{\outket}[2]{|  #1  \rangle  \langle   #2 |}

\newcommand{\D}{{\form{ d}}}

\newcommand{\psif}{\psi_f}

\newcommand{\phs}{\eta}

\newcommand{\form}[1]{{\bm{  #1}}}
\newcommand{\vect}[1]{{\bm{ #1}}}

\newcommand{\rref}[1]{~(\ref{#1})}
\newcommand{\ccite}[1]{~\cite{#1}}

\newcommand{\CC}[1]{{\overline{#1}}}
\begin{document}


\title[Phase/Modulo Relations]{ Geometric Phase and Modulo Relations for Probability  Amplitudes as Functions on Complex Parameter Spaces  }

\author{ Alonso Botero }

\altaffiliation[Currently at: ]{Departamento de F\'{\i}sica, Universidad de los Andes, Apartado Aereo 4976, Bogot\'a, Colombia}
\email{abotero@uniandes.edu.co}

\affiliation{
    Department of Physics and Astronomy,
    University of South Carolina,
    Columbia, SC, 29208 }

\affiliation{
    Centro Internacional de F\'{\i}sica, Ciudad Universitaria, Bogot\'a, Colombia
    }

\date{November 30, 2001}

\begin{abstract}
\bigskip
We investigate general differential relations connecting the respective behaviors of the phase and  modulo of probability amplitudes  of the form $\amp{\psi_f}{\psi}$, where $\ket{\psi_f}$ is a fixed state in Hilbert space and $\ket{\psi}$ is a section of a holomorphic line bundle over some complex parameter space. Amplitude functions on such bundles, while not  strictly holomorphic, nevertheless satisfy  generalized Cauchy-Riemann conditions involving the $U(1)$ Berry-Simon connection on the parameter space. These conditions  entail invertible relations between the gradients of the phase and modulo, therefore allowing for the reconstruction of the phase from the modulo (or vice-versa) and other conditions on the behavior of either polar component of the amplitude.   As a special case, we consider amplitude functions valued on the space of pure states,  the ray space  ${\cal R} = {\mathbb C}P^n$, where  transition probabilities have a geometric interpretation in terms of geodesic distances as measured with the Fubini-Study metric. In conjunction with the generalized Cauchy-Riemann conditions, this geodesic interpretation leads to additional relations, in particular a novel connection between the modulus of the amplitude and the phase gradient, somewhat  reminiscent  of  the WKB formula. Finally, a connection with geometric phases is established. 

\end{abstract}
\maketitle
\section{Introduction}

The study of correlations between the behavior of the phase and modulo of complex  probability amplitudes  is a relevant topic in a number of physical problems such as the   ``phase problem" in diffraction theory   \ccite{phaseret}, the study of phase singularities \ccite{phasesing} and the semi-classical or WKB approximation\ccite{WKBapp} to name a few.   In the phase problem, for instance, the aim is to infer phase information in the diffracted wave from the  observed cross section, which only involves the magnitude of the wave. In the study of phase dislocations, it is known that  regions of vanishing amplitude are characterized by surrounding regions of generally non-uniform vortex-type phase singularities. Finally, in the semi-classical approximation, the phase gradient is in correspondence with the classical momentum and the behavior of the magnitude of the amplitude is correlated to the phase gradient by Liouville's theorem.

From a different standpoint, significant insight into the geometrical meaning of both the modulus and the phase of  probability amplitudes has emerged from the study of the ray space ${\cal R}$ (also known as projective Hilbert space), particularly  in  connection with  geometric phases\ccite{berry,simon,anaha87,sambhand}, quantum information theory\cite{wootters,ravietal}, and other topics falling under the general category of `geometric quantum mechanics'\ccite{anaha90b,brodyhugh}. From the work of Berry\ccite{berry}, Simon\ccite{simon}, Aharonov and Anandan\ccite{anaha87}, it is known that under cyclic evolution a geometric phase factor is acquired by the amplitude,  which is interpreted as the holonomy associated with a natural  connection (the so-called Berry-Simon connection) on the $U(1)$ bundle over ${\cal R}$,  and which is proportional to the symplectic area enclosed by the circuit in ${\cal R}$. Samuel and Bhandari \cite{sambhand} have also shown that the so-called Pancharatnam phase difference between any two states can be expressed  as a line integral of the Berry-Simon connection along the geodesic connecting the two states, as measured with the Fubini-Study metric, the natural metric on ${\cal R}$. Finally, there exists  a natural geometric interpretation  to  transition probabilities in the ray space as the cosine of the geodesic distance with respect to the Fubini-Study metric\cite{anaha90b}, a measure that  is intimately related to information-theoretic measures of statistical distance between two probability distributions \cite{wootters,ravietal}.

In the present paper, the aim is to shed additional insight into the correlation between the phase and magnitude of transition probability amplitudes  from the point of view of  geometric quantum mechanics. Specifically, we study amplitudes of the form $\amp{\psi_f}{\psi}$, where $\ket{\psi_f}$ is any fixed state in Hibert space and $\ket{\psi}$ is parameterized on a complex parameter subspace $\mathcal M$ of the ray space ${\mathcal R}$, or, in particular, the ray space itself. We then obtain  general geometric relations  between the two polar components of the amplitude   arising  from  holomorphicity and metric constraints  natural to such complex parameter spaces. We note that   a number of state  families of broad physical interest are valued on complex parameter spaces, including  the family of coherent or more generally squeezed states, the Bloch sphere of spin-$1/2$ states, as well as complex extensions of real parameter families.

A brief summary of the main results and the structure of the paper is in order. In section \ref{setting} we spell out in greater detail the geometric setting involved, which is more precisely that of {\em holomorphic line bundles} over the complex parameter space ${\cal M}$. Such bundles share with the more general line bundles over arbitrary parameter spaces (arising, for instance, in connection with Berry phases) two important geometric objects, namely the Berry-Simon connection $\form{A} = - i \amp{\psi}{\D \psi}$ and the quantum geometric tensor $\form{H} \propto \mel{\D\psi}{\otimes}{\D \psi}-\amp{\psi}{\D \psi}\otimes\amp{\D \psi}{ \psi}$. The symmetric part of $\form{H}$   gives rise to a  ``quantum" metric on ${\cal M}$, (the Fubini-Study metric when $\mathcal M = \mathcal R$), while the anti-symmetric part, here denoted by $\Omega$, is proportional to the field-strength tensor associated with the connection. There are, however, additional constraints that follow from the fact that ${\cal M}$ is a complex submanifold of ${\cal R}$. In particular, state sections of the corresponding line bundle satisfy  generalized holomorphicity conditions and the base manifold inherits from the ray space its K\"ahler structure. These constraints are then used in  section \ref{relcomp} to show that the polar components of $\amp{\psi_f}{\psi} = \sqrt{p}\, e^{i \phs}$ satisfy a generalized version of  Cauchy-Riemann conditions on the logarithm of $\amp{\psi_f}{\psi}$, the relations
\begin{eqnarray*}
\nabla \log \sqrt{p}  & = &  \ \ \form{\Omega}\cdot\left( \nabla \eta - \form{A} \right) \\
\left( \nabla \eta - \form{A} \right)  & =   &- \form{\Omega}\cdot
\nabla \log \sqrt{p}
\end{eqnarray*}
where the inner product is with respect to the quantum metric on ${\cal M}$. With the aid of these conditions, it is then possible to reconstruct either polar component of the amplitude from the parametric dependence of the other, as well as to obtain additional constraints on the behavior of $p$ and $\eta$.  A brief illustration of the  the generalized Cauchy-Riemann conditions on the Bloch sphere is given in section \ref{blochsph}.  In section \ref{raysp} we turn to the case when $\cal M = \cal R$,  where we explore the consequences of previously obtained results in conjunction with an additional geometric relation that exists between the transition probability $p$ and geodesic distances as measured by the Fubini-Study metric. In particular, we give a generalization of the Samuel and Bhandari result for the  Pancharatnam phase for non-geodesic paths. More importantly, it is shown that  the transition amplitude can be parameterized  entirely in terms of its phase according to the formula 
\begin{equation}
\amp{\psif}{\psi}= \frac{ e^{i \phs} }{ \sqrt{1 + q\| \nabla \eta - \form{A} \|^2}} \, ,
\end{equation}
where $q$ is an arbitrary parameter in the definition of the metric. Prompted by a certain resemblance to the WKB formula $\psi_{WKB}(x) =  e^{i \phs(x)} / \sqrt{|\nabla \eta|}$,  a trajectory interpretation to the phase gradient on ${\cal R}$ is obtained. Finally, in section \ref{geoph}, we establish a connection between our results and the geometric phase acquired during cyclic and non-cyclic evolutions.

\section{Geometry of Holomorphic Line Bundles}
\label{setting}

We devote some time to introduce the relevant geometric aspects that are involved. Let the map $\tilde{\psi}:{\cal M}\rightarrow {\cal H}$ define a family of unnormalized state vectors $\ket{\,\tilde{\psi}(z)\,} \in {\cal H}$, which only depend on a set of local holomorphic coordinates $z^a$ on ${\cal M}$.  The family $\ket{\psi}$ is then obtained by projecting $\ket{\tilde{\psi}}$ onto the set of pure normalized state vectors according to
\begin{equation}\label{defparam}
\ket{\psi(z,\bar{z}) } = \frac{e^{i \gamma(z,\bar{z})}}{\sqrt{\amp{\tilde{\psi}(\CC{z})}{\tilde{\psi}(z)}}}
\ket{\tilde{\psi}(z)}\, ,
\end{equation}
where $\gamma$ is some  (real) phase factor that for the moment will be assumed to be an arbitrary function of $z$ and $\CC{z}$. 

It will also be convenient to keep in mind  alternative parameterizations of $\ket{\psi}$ in terms of the set of real coordinates $(x^a, y^a)$  related to $z^a$ ($ \CC{z}^a$) as usual by $z^a = x^a + i y^a$ ($\CC{z}^a = x^a - i y^a$ ), and more generally in terms of arbitrary real coordinates on ${\cal M}$ which will be denoted by $\xi^\mu$ with the index $\mu$ ranging form $1$ to $2 k $ (throughout the section we use Latin indices $a,b...$ (ranging from $1$ to $k$) to denote complex coordinates or their real and imaginary components and Greek indices to denote general coordinates).

Neglecting for the moment the fact that ${\cal M}$ is a complex manifold, we see that there is a correspondence between a point in  ${\cal M}$, and a pure-state density matrix $|\psi\rangle \langle \psi|$, and therefore a point in the ray space ${\cal R}$, the  equivalence class of states  under the equivalence relation $\ket{\psi} \sim e^{i \phi} \ket{\psi}$. The geometric setting is therefore that of the  $U(1)$ or line  bundle $P({\cal M}, U(1))$ over the parameter space ${\cal M}$\cite{simon,nakahara}, on which a choice of $\ket{\psi}$ with a given phase factor $\gamma$ corresponds to a particular choice of local section.

Now, as is well known in the context of geometric phases\cite{GeoPhaseLet}, there is a natural geometric connection that can be defined on the line bundle over a parameter space ${\cal M}$, which is expressed locally by the so-called Berry-Simon (BS) connection $1$-form $\form{A} = A_\mu \D \xi^\mu$, with components 
\begin{equation}\label{defa}
A_\mu = - i \amp{\psi}{ \partial_\mu \psi} \, 
\end{equation}
where $\partial_\mu = {\partial \over \partial \xi^\mu}$ in arbitrary coordinates.   This connection is naturally induced by the Dirac inner product on Hilbert space $\amp{\phi}{\psi}$   in the sense that the horizontal motion defined by this connection corresponds to infinitesimal variations     orthogonal  to $|\psi \rangle$, i.e., $\amp{\psi}{\delta_{{\rm Horiz}} \psi}=0 $. The resulting covariant derivative of a section $\ket{\psi}$,
\begin{equation}
D_\mu \ket{\psi}\equiv \left[ \partial_\mu - i A_\mu \right] \ket{\psi} \, ,
\end{equation}
therefore satisfies  $\amp{\psi}{D_\mu \psi} = 0$. By virtue of \rref{defparam}, it is clear that under a $U(1)$ gauge transformation $\ket{\psi} \rightarrow e^{i \delta \gamma}\ket{\psi}$, $A_\mu$ transforms as $A_\mu \rightarrow A_\mu + \partial_\mu \delta \gamma$, in such a way that $D_\mu \ket{\psi}$ transforms homogeneously as $D_\mu \ket{\psi} \rightarrow e^{i \delta\gamma}D
_\mu\ket{\psi}$. Furthermore, a $U(1)$ gauge transformation may always be introduced so that the connection form is set to zero at least at one point in ${\cal M}$. As usual, the failure of the covariant derivative to commute in different directions is measured by the curl of $\form{A}$.

When ${\cal M}$ is a complex manifold as is the case in question, there is added richness brought about by the complex nature of the base space. In particular, it is possible to construct a more refined notion of the line bundle over {\cal M},  namely a  {\em Holomorphic line bundle}\cite{kobayashi,nakahara,bottchern,gsw}. The notion of such bundles rests on a generalization of the concept of a holomorphic function, in the sense that by a suitable gauge transformation it is possible to have a section satisfy, {\em at  a given point }, the standard holomorphic condition ${\partial \over \partial{\CC{z}^a}}\ket{\psi}=0$. 
Let us se how this comes about with the parameterization \rref{defparam}. By construction we have that
the unnormalized vector $\ket{\tilde{\psi}}$ satisfies the holomorphic condition
\begin{equation}
{\partial \over \partial{\CC{z}^a}}\ket{\, \tilde{\psi}(z)\, } = 0 \, ,
\end{equation}
with $\CC{z}^{\alpha}= x^a - i y^a$. It is clear however, that $\ket{\psi}$ is not strictly holomorphic, as the anti-holomorphic coordinates $\CC{z}^a$ appear not  only in the phase factor $\gamma$, but more importantly in the  normalization factor which involves the anti-holomorphic map
$\bra{\tilde{\psi}(\CC{z})}$.  Thus we have in general that
\begin{equation}\label{partpsi}
{\partial \over \partial{\CC{z}^a}}\ket{ \psi} = \left[{\partial \over \partial{\CC{z}^a}}\log\frac{e^{i \gamma}}{\sqrt{\amp{\tilde{\psi}}{\tilde{\psi}}}}\right] \, \ket{ \psi }\, .
\end{equation}
Now, taking  the inner product of this expression with $|\psi\rangle$ itself, we find that
\begin{equation}\label{psipartpsi}
\bra{\psi}{\partial \over \partial{\CC{z}^a}}\ket{ \psi}= \left[{\partial \over \partial{\CC{z}^a}}\log\frac{e^{i \gamma}}{\sqrt{\amp{\tilde{\psi}}{\tilde{\psi}}}}\right]  \, .
\end{equation}
Expressing the BS 1-form in the complex basis as 
$
\form{A} =  A_{a}\D z^{a} + A_{\CC{a}}\D \CC{z}^a 
$ with 
$$ A_{a} = -i \amp{\psi}{\partial_a \psi}\, \ \ \ \ A_{\CC{a}} = -i \amp{\psi}{\partial_\CC{a} \psi}
$$
 (where $\partial_a = {\partial \over \partial z^a} $, $\partial_\CC{a} = {\partial \over \partial \CC{z}^a}$),  and  splitting  the covariant derivative $D$ into holomorphic and anti-holomorphic components,  we have
$$ D_a = {\partial \over \partial z^a}  - i A_a\, , \ \ \ \ D_\CC{a} = {\partial \over \partial \CC{z}^a} - i A_\CC{a}\, .
$$ 
Thus, we find from \rref{partpsi} and \rref{psipartpsi} that the section $\ket{\psi}$ satisfies  a generalized ``gauge covariant" holomorphic condition \begin{equation}\label{genholocond2}
D_\CC{a}\ket{\psi} \, = 0 .
\end{equation}
However, it is always possible to gauge away the BS connection  at least at one point. At that point then, the section satisfies the usual holomorphic condition $\partial_{\CC{a}}\ket{\psi} = 0$. Thus,  modulo a $U(1)$ gauge transformation,  $\ket{\psi}$ is a  locally holomorphic section. For future reference, we shall also need the dual, now anti-holomorphic condition, on the bra $\bra{\psi}$. This is given by
\begin{equation}
\bra{D_{a}\psi}  = \left[ \partial_a + i A_a \right]\bra{\psi} =0 \, .
\end{equation}

We now consider geometric aspects of the base space ${\cal M}$ and introduce additional objects that will be of use later. Viewed as a general parameter space   ${\cal M}$, the horizontal motion associated with  the BS connection on the line bundle over ${\cal M}$ induces naturally on the  base space ${\cal M}$ a gauge-invariant rank-2 hermitian tensor 
\begin{equation}
H_{\mu \nu} =  \, q\, \amp{D_\mu \psi }{D_\nu \psi}  \, ,
\end{equation}
 which Berry \cite{WilcShap} has named the {\em quantum geometric tensor}. Here, $q$ is any  strictly positive real number to be adjusted for convenience. The real part of $H_{\mu \nu}$ is positive definite and symmetric, and thus defines a metric $g_{\mu \nu}$ on ${\cal M}$, the {\em quantum metric}, with line element
\begin{equation}\label{quantummet}
ds^2 = g_{\mu \nu}d\xi^\mu d\xi^\nu  = q \left[ \, \amp{d \psi }{d \psi} -\amp{d \psi }{ \psi}\amp{ \psi }{d \psi} \, \right]\, .
\end{equation}
In turn, the imaginary part of $H$ is anti-symmetric and is closely related to the curl of the BS connection 1-form $\form{A}$:
\begin{equation}
\form{\Omega} = {\rm Im}H =  \frac{q}{2}\bra{\D\psi }\wedge\ket{\D \psi}= \frac{q}{2}\D \form{A}\, .
\end{equation}
Since $\D^2 =0$ it follows that $\form{\Omega}$ is automatically closed. 

When ${\cal M}$ is the base space for the hermitian line bundle, considerable simplifications follow. First of all, from the generalized holomorphic condition
$\ket{D_{\CC{a}}\psi}=0$ and its dual, we  have that in complex coordinates the quantum geometric tensor takes as components
\begin{equation}
H_{\CC{a} b} = \amp{D_\CC{a} \psi }{D_b \psi}\, \ \ \ \  H_{a \CC{b}} = 0 \, .
\end{equation}
This implies that the metric, as well as the 2-form $\Omega$ may be written out as
\begin{eqnarray}\label{metsimpcomp}
\form{g} & = & g_{a \CC{b}}\, \D z^a \otimes \D \CC{z}^b + g_{\CC{a} b}\, \D \CC{z}^a \otimes \D z^b \, \nonumber \\
\form{\Omega} & = & i g_{a \CC{b}}\, \D z^a \wedge \D \CC{z}^b \, .
\end{eqnarray}
where 
$
g_{a \CC{b}}= g_{ \CC{b}a } = \frac{1}{2}H_{a \CC{b}}\,  .
$
Note that if the metric is non-degenerate as we shall assume henceforth, it then follows, on the one hand, that  the $U(1)$ connection  $\form{A}$ is non-trivial, and on the other, that both $\form{g}$ and $\form{\Omega}$ admit inverses. In particular the inverse metric takes the form
\begin{equation}
\form{g}^{-1} = g^{a \CC{b}} \partial_a \otimes \partial_\CC{b} + g^{ \CC{a} b} \partial_\CC{a} \otimes \partial_b 
\end{equation}
where  $g^{ \CC{a} b} = g^{b \CC{a}} $ satisfies $g_{a \CC{b}}g^{\CC{b} c} = \delta_{a}^{c}$.

To understand the significance of \rref{metsimpcomp}, we now introduce the so-called complex structure, the defining tensorial object for  a complex manifold. In complex coordinates, the complex structure  tensor $J$ takes the canonical form
\begin{equation}\label{compstruct}
J^{a}{}_b = i\, \delta^{a}{}_b \, \ \ \ \  J^{\CC{a}}{}_\CC{b}  =- i\, \delta^{\CC{a}}{}_\CC{b}
\end{equation}
with the remaining components vanishing. The complex structure satisfies $J^\mu{}_\lambda J^\lambda{}_{\nu} = -\delta^\mu{}_\nu$ (i.e., $J^2 = -1$) and implements the multiplication by $i$ ($-i$) on vector fields with holomorphic (anti-holomorphic) indices.  In terms of $J$, it is readily verified that the metric satisfies  
\begin{equation}\label{hermcond}
g_{\mu \nu} = J^{\gamma}{}_{\mu} J^{\lambda}{}_\nu g_{\gamma \lambda} \, .
\end{equation}  In this case on says that the metric  is  {\em Hermitian}. In turn, the two-form $\form{\Omega}$ is what is known as the {\em K\"ahler form} of the metric, defined by 
\begin{equation}
\Omega_{\mu \nu} = g_{ \lambda \nu} J^{\lambda}{}_{ \mu } \,  ,
\end{equation} 
i.e., $\Omega_{\mu \nu}= - J_{\mu \nu}$.  The expressions for $\form{g}$ and $\form{\Omega}$ in \rref{metsimpcomp}, where $g_{a b}=g_{\CC{a}\CC{b}}= \Omega_{a b} = \Omega_{\CC{a}\CC{b}}=0$,  are the canonical forms  that a Hermitian metric and its K\"ahler form take in complex coordinates. 

When the K\"ahler form $\Omega$ is closed, as in our case,    ${\cal M}$ is   known as  a K\"ahler manifold and the metric a K\"ahler metric. The offshoot of this is a compatibility between the Hemitian and Riemannian structures of the manifold, embodied by the fact that 
\begin{equation}
\D \Omega =0 \Leftrightarrow \nabla_\mu J^{\nu}{}_{\lambda} =  0 \, ,
\end{equation}
where $\nabla_\mu$ denotes covariant covariant differentiation of ordinary tensor fields on ${\cal M}$ with respect to the affine connection associated with the metric $g$.  The condition $d \form{\Omega}=0$ entails that the hermitian components $g_{a \CC{b}}$ of the metric and the K\"ahler form satisfy in complex coordinates the symmetry conditions:
\begin{equation}\label{kahlmet}
\partial_c g_{a \CC{b}}  = \partial_a g_{c \CC{b}}\, \ \ \ \ \  \partial_\CC{c}  g_{\CC{a} {b}}  =\partial z^\CC{a} \partial g_{\CC{c} {b}} \, .
\end{equation}
From the definition of the affine connection in arbitrary  coordinates, $ \Gamma^{\mu}_{\nu \lambda} = \frac{1}{2}g^{\mu \gamma}(\partial_\mu g_{\nu \gamma} + \partial_{\nu} g_{\mu \gamma} - \partial_{\gamma} g_{\mu \nu})$, it is then straightforward to verify that in complex coordinates the affine connection  takes the form
\begin{equation}\label{cristofherm}
\Gamma^a{}_{b c} = g^{ \CC{d}a}\partial_{b} g_{c \CC{d}} \, \ \ \ \ \Gamma^\CC{a}{}_{\CC{b} \CC{c}} = g^{\CC{a} d }\partial_{\CC{b}} g_{ d \CC{c}}\, ,
\end{equation}
with the symbols mixing   holomorphic and ani-holomorphic indices vanishing. Covariant differentiation with respect to a holomorphic (anti-holomorphic) coordinate therefore acts like regular differentiation on  anti-holomorphic (holomorphic) indices. Another way of saying this is that the {\em affine} connection preserves the separation between holomorphic and anti-holomorphic tensor fields.  We  remark that in complex manifolds that are not K\"ahler, it is still possible to define a Hermitian connection taking the form \rref{cristofherm} and satisfying $\nabla J =0$, but this connection will not coincide with the affine connection. 

A second consequence of the symmetry conditions\rref{kahlmet} is that the K\"ahler metric  may be derived locally from a scalar potential function, the so called {\em K\"ahler potential}, according to
$
g_{a \CC{b}} = \partial_a \partial_{\CC{b}} K(z,\CC{z})  \, .
$
This can be seen by noting from \rref{psipartpsi} that  
\begin{equation}
\form{A} = \D \gamma + \frac{1}{2 i}\partial_{a}\, \log \amp{\tilde{\psi}}{\tilde{\psi}}\, \D z^a - \frac{1}{2 i}\partial_{\CC{a}}\, \log \amp{\tilde{\psi}}{\tilde{\psi}} \D \CC{z}^a \, ,
\end{equation}
from which we see that
\begin{equation}
\form{\Omega} = \frac{q}{2}\D \form{A}= \frac{i q}{2} \partial_a \partial_{\CC{b}} \log \amp{\tilde{\psi}}{\tilde{\psi}}\D z^a \wedge \D \CC{z}^b \, .
\end{equation}
Consequently, from  \rref{metsimpcomp}, we have that
\begin{equation}
g_{a\CC{b}} = \frac{q}{2}\partial_a \partial_{\CC{b}}\log \amp{\tilde{\psi}}{\tilde{\psi}} \, .
\end{equation}
so that an appropriate {\em K\"ahler} potential  is given by
\begin{equation}
\tilde{K} = \frac{q}{2} \log \amp{\tilde{\psi}}{\tilde{\psi}} \, .
\end{equation}
Note however that this potential is not uniquely defined, since one is free to add to $K(z,\CC{z})$ any function of the form  $f_1(z) + f_2(\CC{z})$ without changing $g_{a \CC{b}}$. Note finally that  a choice of gauge in which $\gamma(z,\bar{z})= Re[f(z)] = \frac{1}{2}\left[ f(z)+f^*(\bar{z}) \right]$ for some arbitrary holomorphic function $f(z)$, is equivalent to a re-parameterization of $|\psi \rangle$ in which in \ref{defparam}, the phase $\gamma$ is set to zero and the unnormalized vector $\ket{\tilde{\psi}}$ gets replaced by $\ket{\tilde{\psi}'} = e^{f(z)}\ket{\tilde{\psi}}$. In such case, all geometric objects of interest for us, namely, the  Berry-Simon connection and the second-rank tensors obtained from the quantum geometric tensor  can be derived from the K\"ahler potential $\tilde{K'} = \frac{q}{2} \log \amp{\tilde{\psi}'}{\tilde{\psi}'} \, $. The freedom that remains in the choice of $f$ corresponds to the freedom in the definition of the K\"ahler potential.

\section{Generalized Cauchy-Riemann Conditions}
\label{relcomp}
\bigskip

We now explore the consequences on the behavior of the polar components of any transition amplitude
$$
\amp{\psi_f}{\psi(\xi)}= \sqrt{p(\xi)}\,  e^{i \eta(\xi)} \, ,
$$ where $\ket{\psi(\xi)}$ is a section of a holomorphic line bundle over a complex submanifold ${\cal M \in \cal R}$, i.e., of the form \rref{defparam}, as described in the previous section, and $\xi$ are arbitrary real coordinates on ${\cal M}$.  

The first thing to note is that  due to the arbitrariness in the definition of the phase $\gamma$ in \rref{partpsi}, the notion of phase  $\amp{\psi_f}{\psi}$ is tied to the choice of gauge. Specifically, the phase 
$ \eta \equiv \arg{\amp{\psi_f}{\psi}} $,  transforms under the $U(1)$ gauge transformations  $\ket{\psi} \rightarrow e^{ i \delta \gamma} \ket{\psi}$ as
$$ \eta \rightarrow \eta + \delta\gamma .$$   It then becomes convenient  to introduce a gauge-  invariant notion of  phase variation by means of the $BS$-connection This is done by  defining a {\em gauge invariant phase gradient}
\begin{equation}
V_\mu \equiv \partial_\mu \eta - A_\mu \, . 
\end{equation}
Clearly, the 1-form $\form{V} = V_\mu \D \xi^\mu$ is not closed but rather satisfies
$\D \form{V} = - \D \form{A}- \frac{2}{q} \form{\Omega}$. The modulus of $\amp{\psi_f}{\psi}$,
$
\sqrt{p} \equiv |{\amp{\psi_f}{\psi}} | $, 
is of course gauge invariant. 

Now, since  $\ket{\psi_f}$ is assumed to be a constant vector, it follows from \rref{genholocond2} that the amplitude $\amp{\psi_f}{\psi}$ is as well subject to the generalized holomorphic (anti-holomorphic) conditions  when expressed in local complex coordinates
\begin{eqnarray}\label{ampholo}
\left[{{\partial \over \partial\CC{z}^a} - i A_{\CC{a}}}\right]\amp{\psif}{\psi} & = & 0\, , \\
\left[{{\partial \over \partial z^a} + i A_a }\right]\amp{\psi}{\psif} & = & 0 \, .
\end{eqnarray}
Assuming then that $\amp{\psif}{ \psi} \neq 0$, the logarithm of the amplitude can be  defined analytically, and we find that
\begin{eqnarray}\label{kasdpanti}
i \left[ \partial_\CC{a}\eta - A_\CC{a}\right] + \partial_{\CC{a}}\log \sqrt{p}  & = & 0 \nonumber \\
-i \left[ \partial_a\eta - A_a\right] + \partial_{a}\log \sqrt{p}  & = & 0 \, .
\end{eqnarray}
As mentioned earlier, by a suitable choice of gauge it is possible to have the section $\ket{\psi}$ satsify ordinary Cauchy-Riemann conditions at a specified point. Correspondingly,   the above conditions can be brought  locally to the form of ordinary Cauchy-Riemamann conditions. 

For our purposes, it will be more convenient to cast the above expression in terms of the K\"ahler form $\form{\Omega}$ which has a more immediate interpretation in terms of the  Berry-Simon connection $\form{A}$ (recall that $\form{\Omega} = \frac{q}{2} \D \form{A}$). Using the facts that $ J^\nu{}_{\mu}V_\nu = J_\nu{}_{\mu}V^\nu =  \Omega_{\mu \nu}V^\nu = \Omega_{\mu}{}^{ \nu}V_\nu$ and that in mixed-rank form $J^2 = \Omega^2 =-1$, we then have the following alternative expressions:
\begin{subequations}
\label{cauchgen}
\begin{eqnarray}
\partial_\mu \log \sqrt{p} & = & \Omega_{\mu}{}^{\nu} \left[ \, \partial_\nu \eta - A_\nu \right]\, ,\label{cauchgen1} \\
\partial_\mu \eta  & = & A_\mu - \Omega_{\mu}{}^{\nu}\partial_\nu\log \sqrt{p} \label{cauchgen2}\, .
\end{eqnarray}
\end{subequations}
We shall refer to these as the {\em generalized Cauchy Riemann conditions} satisfied by the polar components of the amplitude $\amp{\psi_f}{\psi}$. These  conditions  constitute the first important result of the paper, and will serve as a starting point for a number of additional relations that will be derived in the forthcoming. 

The most important consequence of \rref{cauchgen}  is the existence of the reconstruction formulas on ${\cal M}$ 
\begin{equation}
\eta(\xi) -\eta(\xi_o) =  \int_{\xi_o}^{\xi}\, d\xi^\mu  \left[A_\mu -\Omega_\mu{}^{\nu} \partial_\nu \log\sqrt{p}\, \right] 
\end{equation}
and
\begin{equation}
\sqrt{p(\xi)\over p(\xi_o)} = \exp\left[{\int_{\xi_o}^{\xi} \, d\xi^\mu \Omega_\mu{}^{\nu}\left[ \, \partial_\nu \eta - A_\nu \right] }\right] \, ,
\end{equation}
by means of which one polar component of the amplitude $\amp{\psif}{ \psi }$ can be obtained from the other by line integration once  the connection is specified. Since both formulas arise from exact differentials, the choice of  integration path can be left arbitrary as long as any two paths may be deformed continuously into one another within a simply-connected region  excluding singularities. However, since the individual terms in the integrands are not generally  exact, all terms  must be evaluated along the same path.  This path independence must reflect itself, therefore, in  ancillary relations that both $p$ and $\eta$ have to satisfy in order to guarantee that the left hand sides of \rref{cauchgen1} and \rref{cauchgen2} are exact differentials, conditions that will be examined in more  detail shortly. 

Before doing so, we  use the fact that the quantum  metric\rref{quantummet} on ${\cal M}$ is a hermitain metric to establish relations on the magnitude and angle between the gauge invariant gradients $\nabla \eta - \form{A}$ and $\nabla \log \sqrt{p}\,$ on ${\cal M}$.   The hermitian condition on the metric is that the complex structure should preserve the inner product, i.e. $\form{X} \cdot \form{Y} = (J \form{X}) \cdot (J \form{Y})$. This implies from the generalized Cauchy-Riemann conditions that the gauge invariant gradients have the same magnitude as defined in terms of the quantum metric
\begin{equation}\label{equalnorm}
|\nabla \log \sqrt{p}\, | = |\nabla \eta - \form{A}| \, .
\end{equation}
 A second property of a hermitian metric is that $\form{X} \cdot \form{J X} = 0$, a property that in a K\"ahler manifold follows automatically from the anti-symmetry of the  K\"ahler form $\form{\Omega}$. It follows therefore from \rref{cauchgen} that
\begin{equation}\label{orthograds}
\nabla p \, \cdot \left(\nabla \eta - \form{A}\right) = 0 \, .
\end{equation}
A particular consequence is therefore that  lines of constant phase and constant transition probability necessarily meet at right angles wherever $\vect{A}$ is made to to vanish by a choice of gauge.

Next, we turn to the previously mentioned integrability conditions.  The most evident one comes from re-arranging
\rref{cauchgen2} to read
$
\Omega_{\mu}{}^{\nu}\partial_\nu\log \sqrt{p}   = A_\mu - \partial_\mu \eta \, ,
$
in which case we see that the one-form $\form{\Omega} \cdot \D \log \sqrt{p}$ is equivalent to the BS connection up to the gauge term $\D \eta$. Correspondingly, the curl of  $\Omega_{\mu}{}^{\nu}\partial_\nu\log \sqrt{p}$ must lead to the same curl of $A_\mu$, which, up to a constant, is nothing more  than the K\"ahler form. Using the fact that the K\"ahler form is covariantly constant, we then have 
$$
\Omega_{\nu}{}^{\gamma}\nabla_\mu\nabla_\gamma\log \sqrt{p} - \Omega_{\mu}{}^{\gamma}\nabla_\nu\nabla_\gamma\log \sqrt{p} = \frac{2}{q}\Omega_{\mu \nu} \, .
$$
Multiplying on both sides by the K\"ahler form and using $\Omega^2 = -1$,  the equation can then be transformed to  
\begin{equation}\label{intcond1}
\left[\nabla_\mu\nabla_\nu+ \Omega_{\mu}{}^{\alpha}\Omega_{\nu}{}^{\beta}\nabla_\alpha\nabla_\beta \right]\log \sqrt{p} = -\frac{2}{q}g_{\mu \nu} \, .
\end{equation}
The interpretation of this equation becomes more straightforward in complex  coordinates, in which case it reads
\begin{equation}
q \nabla_a \nabla_\CC{b}\log\sqrt{p} = - g_{a \CC{b}} \, .
\end{equation}
Since $\nabla_a \nabla_\CC{b}\log\sqrt{p} = \partial_a \partial_\CC{b}\log\sqrt{p}$, we further see that the condition {\em is  that  $-q\log\sqrt{p}$ is a K\"ahler potential for the quantum metric on ${\cal M}$}. This can be seen more clearly by noting from \rref{defparam} that
\begin{eqnarray*}
q\log \sqrt{p} & = & \frac{q}{2}\log \left[ \amp{\psi_f}{\psi}{\amp{\psi}{\psif}} \right]\, \\ & = & -\frac{q}{2} \log \amp{\tilde{\psi}}{\tilde{\psi}} +  q\log \amp{\psi_f}{\tilde{\psi}}+ q
\log \amp{\tilde{\psi}}{\psif}
 \, .
\end{eqnarray*}
Apart from the purely holomorphic and anti-holomorphic functions $\log \amp{\psi_f}{\tilde{\psi}}$ and $\log \amp{\tilde{\psi}}{\psif}$ respectively, this is nothing more than minus the K\"ahler potential $\tilde{K} = \frac{q}{2}\log \amp{\tilde{\psi}}{\tilde{\psi}}$ mentioned in Section \ref{setting}.

Concerning the second integrability condition, we can use the fact that $\partial_\mu \log \sqrt{p}$ is a gradient to obtain from \ref{cauchgen1} that
\begin{equation}\label{intcond2}
\left[\nabla_\mu \delta_\nu^\beta+ \Omega_{\mu}{}^{\alpha}\Omega_{\nu}{}^{\beta}\nabla_\alpha \right](\nabla_\beta\eta - A_\beta)=0\, ,
\end{equation}
a condition that in complex coordinates takes the form
\begin{equation}
\nabla_a( \nabla_\CC{b}\eta - A_\CC{b}) + \nabla_\CC{b}( \nabla_a \eta - A_a) = 0 \, .
\end{equation}
Further insight into this condition is obtained  from the expression \rref{} in which case we see that $\nabla_a A_\CC{b} + \nabla_\CC{b} A_a $ is nothing more than $2 \partial_{a}\partial_\CC{b}\gamma$. Therefore, under the restricted choice of $\gamma = \frac{1}{2}\left[f(z) + \bar{f}(\bar{z})\right]$ mentioned previously as the condition in which all geometric quantities can be derived from the Ka\"hler potential, we have that $\partial_{a}\partial_\CC{b}\gamma =0$ and therefore that $$\nabla_a \nabla_\CC{b}\eta  = 0 \, .$$ In other words,  the second condition expresses the fact that  modulo a gauge term mixing holomorphic and anti-holomorphic coordinates, the phase $\eta$ is a linear combination of a purely holomorphic and a purely anti-holomorphic function. This can be seen most clearly by noting from \rref{defparam} that
$$
\eta = \gamma + \frac{1}{2 i} \log \amp{\psi_f}{\tilde{\psi}}- \frac{1}{2 i}
\log \amp{\tilde{\psi}}{\psif} \, .
$$

We see therefore that the integrability conditions \rref{intcond1} and \rref{intcond2} ensuring the consistency of the generalized Cauchy-Riemann conditions \ref{cauchgen} are  rather trivial consequences of  the  parameterization \rref{defparam} of state sections on a holomorphic line bundle. Still, they lead to non-trivial constraints on the behavior of the phase and modulo of the amplitude when the parameter space is viewed as a general Riemannian manifold. 

In particular,  by contracting indices in \rref{intcond1} and \rref{intcond2}\,   we determine that the phase and modulo of $\amp{\psi_f}{\psi}$ satisfy locally the scalar conditions
\begin{subequations}
\label{laps}
\begin{eqnarray}
\nabla\cdot(\nabla \eta - \form{A}) & = & 0 \label{laps1}\\
\nabla ^2 \log \sqrt{p}  & = & -\frac{2 k}{q}\label{laps2}  \, ,
\end{eqnarray}
\end{subequations}
where $k$ is the complex dimension of ${\cal M}$ and $\nabla \cdot$ and $\nabla^2 $ are the divergence and Laplacian operators on ${\cal M}$ associated with the quantum metric.  Note that since it is always possible to choose $\nabla \cdot \form{A} =0$ (for instance with the restricted choice of gauge mentioned previously), the first condition can always be brought to the form $\nabla^2 \eta = 0$. 

Finally, it is interesting to note that from the scalar conditions \rref{laps} and the relations \rref{equalnorm} and \rref{orthograds} one obtains
\begin{eqnarray}
\nabla\cdot\left[ p\, (\nabla \eta - \form{A})\right] & = & 0 \\
\frac{1}{2}|\nabla \eta - \form{A}|^2 - \frac{1}{2}\frac{\nabla ^2 \sqrt{p}}{\sqrt{p}}  & = & \frac{ k}{q}\, ,
\end{eqnarray}
a set of equations analogous to the generalized Hamilton-Jacobi equation and the probability conservation equation arising from the time-independent Schr\"odinger  equation  a free particle in a magnetic field. That the probability amplitude $\amp{\psi_f}{\psi}$ therefore satisfies on ${\cal M}$ the corresponding Sch\"odinger equation
$$
-\frac{1}{2}(\nabla  -i \form{A})\cdot(\nabla  -i \form{A})\, \amp{\psi_f}{\psi} = \frac{ k}{q}\amp{\psi_f}{\psi}
$$
can be verified by noting that in complex coordinates $g^{a \CC{b}}D_a D_{\CC{b}}\amp{\psi_f}{\psi}=0$ and using the commutation relation $[ D_a , D_\CC{b} ] = -\frac{2 i}{q}\Omega_{a \CC{b}} = \frac{2 }{q}g_{a \CC{b} }$.  The analogy  between $(\nabla \eta - \form{A})$ and a velocity field suggests that it may be possible to establish a trajectory interpretation for the invariant phase gradient. We shall see in Section \ref{raysp} that such an interpretation is indeed possible on the ray space. 

\section{Phase/Modulo relations on the Bloch Sphere}
\label{blochsph}

Let us for the moment flesh out the preceding results with a simple concrete illustration. Consider the family of spin-$1/2$ states $\ket{\hat{n}}$ represented by points on the Bloch sphere labeled by the usual polar angles $\theta, \phi$,
\begin{equation}\label{ketangle}
\ket{\hat{n}} = \left(\begin{array}{c} \cos {\theta \over 2} \\ \sin{\theta \over 2} e^{ i \phi} \end{array}\right) \, ,
\end{equation}
where the basis used is the standard $\ket{\pm}$ eigenbasis of $\sigma_3$. As is well known, the two-sphere is in fact a complex manifold, namely the complex projective space $\mathbb{C}P^1$. To see this, note that the parameterization of the unnormalized state $\ket{\tilde{\psi}}$
 \begin{equation}
\ket{\tilde{\psi}(z)} = \left(\begin{array}{c} 1 \\ z \end{array}\right) \, ,
\end{equation}
maps, according to \rref{defparam}, to the quantum state section \rref{ketangle}  after the identification
\begin{equation}
z = \tan {\theta \over 2} \, e^{i \phi}\, , \ \ \ \ \gamma = 0 \, .
\end{equation}
 The map corresponds to a stereographic projection of the sphere to the complex plane, mapping the south pole into $z = \infty$.  

We proceed by calculating the geometric objects of interest. From the Ka\"hler potential, $\tilde{K}= \frac{q}{2}\log \amp{\tilde{\psi}}{\tilde{\psi}} = \frac{q}{2}\log(1 + z\CC{z})$, it is straightforward to compute the metric element, i.e.,
\begin{equation}
ds^2 = q \frac{dz d\CC{z}}{(1 + |z|^2)^2} = \frac{q}{4}\left[ d \theta^2 + \sin^2 \theta \, d \phi^2 \right] \, .
\end{equation}
Choosing $q =4$ for this example, the  quantum metric reduces to the usual metric on the unit sphere, with non vanishing components
$$ 
g_{\theta \theta} = 1 \, \ \ \ \  g_{\phi \phi} = \sin^2 \theta \, .
$$
The BS connection form is more straightforward to calculate from \rref{ketangle} and we find that
\begin{equation}
\form{A} = \sin^2 {\theta \over 2} \, \D \phi = \frac{1}{2} ( 1 - \cos \theta )\,  \D \phi \, .
\end{equation}
The BS connection leads therefore to the K\"ahler form 
\begin{equation}
\form{\Omega} = \frac{q}{2} \D \form{A} \, = \, \sin\theta \, \D \theta  \wedge \D \phi \, ,
\end{equation}
which is immediately recognized as the volume form for the unit two-sphere. 

With this, it is then possible to express the generalized Cauchy Riemann conditions for the polar components of some amplitude $\amp{\psi_f}{\hat{n}}$ in a more conventional form by embedding them in  three dimensional space.  Letting $\hat{n}$ now stand for $\vec{r}/r$ and using standard vector notation, the connection becomes
\begin{equation}\label{diraca}
\vec{A} = \frac{1}{ 2 r} \left[ \frac{1 - \cos \theta}{\sin \theta} \right]\hat{\phi}\, ,
\end{equation}
which is the usual ``Dirac string"  vector potential for a magnetic  charge $1/2$ located at  the origin and with the string singularity along the south pole. Relations \rref{cauchgen2} and \rref{cauchgen2} now read
\begin{eqnarray}
\vec{\nabla}
 \log\sqrt{p} & = &  - \hat{n}  \times \left(\vec{\nabla} \eta - \vec{A} \right)\, \nonumber \\
\vec{\nabla} \eta  & = & \vec{A} + \hat{n}  \times \vec{\nabla} \log\sqrt{p} \, ,
\end{eqnarray}
where $p$ and $\eta$ are assumed to depend only on the polar angles. 
We verify this in a simple example. Take 
\begin{equation}
\amp{-}{\hat{n}} = \sin {\theta \over 2} \, e^{ i \phi} \, ,
\end{equation} 
wherefrom we see that $\sqrt{p} = \sin {\theta \over 2}$ and $\eta = \phi$ so that 
$$
\vec{\nabla} \log\sqrt{p} = \frac{1}{2 r }\cot{\theta \over 2} \hat{\theta}
$$ and $\vec{\nabla} \eta = \frac{1}{ r \sin \theta}\hat{\phi}$. The gauge invariant phase gradient is therefore
$$
\vec{\nabla} \eta - \vec{A} = \frac{1 }{2 r \sin \theta}\left[1 - \frac{1}{2}(1 - \cos\theta)\right]\hat{\phi} = \frac{1 }{2 r}\cot{\theta \over 2}\hat{\phi}
$$
and thus we verify that $\vec{\nabla} \log\sqrt{p} = -\hat{n} \times (\vec{\nabla} \eta - \vec{A})$.

 \begin{figure} \epsfxsize=3truein 
       \centerline{\epsffile{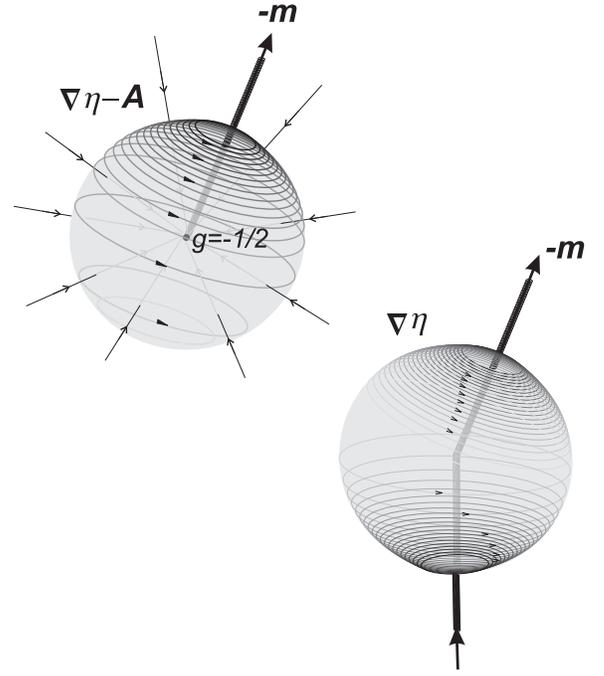}} 
   \caption[]{Interpretation of a) the gauge invariant phase gradient as the vector potential on the unit sphere for a magnetic charge $-1/2$ and b) the phase gradient as the vector potential on the sphere for a trapped flux line of flux $2 \pi$.  } 
     \label{monoflux} \end{figure}

More generally, we obtain a coordinate-independent geometric interpretation of the phase  gradient for a transition amplitude $\amp{\hat{m}}{\hat{n}}$ for fixed $\ket{\hat{m}}$ not necessarily on the same section as $\ket{\hat{n}}$. Since the transition probability is
\begin{equation}
p = |\amp{\hat{m}}{\hat{n}}|^2 =  \frac{1}{2}\left( 1 + \hat{n} \cdot \hat{m}\right) \, ,
\end{equation}
using $\vec{\nabla}(\hat{n}\cdot \hat{m}) = (\hat{m} - (\hat{m}\cdot\hat{n}) \hat{n})/r$  we obtain
\begin{equation}\label{releta}
 \hat{n} \times  \vec{\nabla} \log \sqrt{p}  =  \left( \frac{1}{2 r} \right) \frac{\hat{n}\times \hat{m}}{1 + \hat{n}\cdot\hat{m}} \, .
\end{equation}
But noting that the vector potential \rref{diraca} may also be expressed as
$
\vec{A} = \left( \frac{1}{2 r} \right) \frac{ \hat{z}\times\hat{n}}{1 + \hat{n}\cdot\hat{z}} 
$,
 we see by comparison that $\hat{n} \times  \vec{\nabla} \log \sqrt{p}$, and hence the invariant phase gradient $\vec{\nabla}\eta - \vec{A}$, is the vector potential in a {\em fixed} gauge (specified by \rref{releta}) for a magnetic monopole of charge $-1/2$ centered at the origin with the string singularity running along the $-\hat{m}$ axis.  Thus,  the phase gradient $\vec{\nabla}\eta$ is nothing more than the sum of two vector potentials for two magnetic charges of opposite sign at the origin, with the respective Dirac flux strings running along the directions $-\hat{z}$ and $-\hat{m}$. Equivalently, we can say that the phase gradient $\vec{\nabla} \eta$ is the local vector potential at the exterior of single trapped-flux-line running along  $-\hat{z}$ axis into the origin and exiting along  the $-\hat{m}$ axis (Fig. \ref{monoflux}), with the flux enclosed being $1/2 \times 4 \pi = 2 \pi$. 

Some global properties of the phase dependence now become evident. First,  the invariant phase gradient has only vortex-like  singularity at $\hat{n} = -\hat{m}$, where the amplitude $\amp{\hat{m}}{\hat{n}}$  vanishes, with a circulation $ \oint (\vec{\nabla} \eta - \vec{A}).d\vec{l} = 2 \pi$. On the other hand, the actual phase gradient has, generically, two such singularities with circulation $\oint \vec{\nabla} \eta .d\vec{l} = \pm 2 \pi$ at the two points on the unit sphere where the flux line crosses. One of these points is fixed to be $\hat{n} = -\hat{m}$ corresponding to the actual singularity at $\amp{\hat{m}}{\hat{n}}=0$;  the other point reflects the string singularity in the connection and is therefore dependent on the choice of section $\ket{\hat{n}}$. Note that while additional singularities may be created by means of singular gauge transformations, the string singularity associated with the connection cannot be removed. The exception is when the string singularity happens to be precisely at $\hat{n} = -\hat{m}$, in which case both singular points disappear and  the phase is essentially a constant up to non-singular gauge transformations.

\section{Additional Phase/Modulo Relations on the Ray Space}
\label{raysp}

So far, we have considered phase/modulo relations for transition amplitudes of the form $\amp{\psi_f}{\psi}$ where $\ket{\psi}$ is a section of the holomorphic bundle over an arbitrary complex pure quantum state manifold.  Any such space is  itself a complex submanifold of the so-called ray space ${\cal R}$, the entire space of pure quantum states modulo a phase transformation.  If $n+1$ is the dimensionality of the Hilbert space of the quantum system, then the ray space is the complex projective space ${\mathbb C}P^{n-1}$. A state section $\ket{\psi}$ over ${\cal R}$ is therefore a section over a holomorphic line bundle as well, and hence the results of the previous section hold without change. However, on the ray space, it is possible to establish an additional geometric relation between the transition probability $p = |\amp{\psi_f}{\psi}|^2$ and geodesic distances on ${\cal R}$ as measured with the quantum metric. By virtue of the generalized Cauchy-Riemann conditions \ref{cauchgen}, this new relation has far reaching-consequences, as we now show.

On the ray space, the quantum metric 
$
ds^2 =  q \left[\, \amp{d \psi }{d \psi} -\amp{d \psi }{ \psi}\amp{ \psi }{d \psi} \, \right]\, 
$
is known as the {\em Fubini-Study} metric, and is the most  natural Riemannian metric on the ray space as it the only one invariant under unitary transformations.  Geometrically, the metric arises quite naturally by defining for two arbitrary  rays in Hilbert space $[\,  \ket{\phi}\, ]$ and $[\,  \ket{\psi}\, ]$ (represented by the normalized states $\ket{\phi}$ and $\ket{\psi}$), the distance function \ccite{anaha90b}
\begin{equation}\label{cosrel}
s(\phi,\psi) = \sqrt{q}\, \cos^{-1}|\amp{\psi}{\phi}|\, .
\end{equation}
The Fubini-Study metric is then obtained by choosing $\ket{\phi}$ and $\ket{\psi}$ on the same section and taking the limit when $\ket{\phi}$ goes to $\ket{\psi}$, in which case 
 $$
\ket{\phi} \simeq \ket{\psi} + \ket{d\psi} + \frac{1}{2}\ket{d^2\psi}\, ,
$$ thus yielding the infinitesimal distance  function
$
ds(\phi,\psi) = \sqrt{q}\sqrt{\amp{d \psi }{d \psi} -\amp{d \psi }{ \psi}\amp{ \psi }{d \psi}}\, .
$

From the above considerations it holds, therefore, that the modulus of the amplitude $\amp{\psi_f}{\psi(\xi)}$ can be expressed as a function of the Fubini-Study  geodesic distance $s(\xi)$ between the  rays  $[\,  \ket{\psi}\, ]$ and the fixed state $[\,  \ket{\psi_f}\, ]$, according to \begin{equation}
\label{dist}
 \sqrt{p(\xi)} = \cos \left( \frac{s(\xi)}{\sqrt{q}} \right)  \, .
\end{equation}
From this, we deduce that the gradient of $\log \sqrt{p}$ is given by
$$
\nabla \log \sqrt{p} = -\frac{1}{\sqrt{q}}\tan\left(\frac{s}{\sqrt{q}}\right)\, \nabla s \, . $$
Now, since  the modulus of the gradient measures the rate of change with respect to the metric length,  it is clear that 
\begin{equation}
|\nabla s |^2 = 1 \, .
\end{equation}
Translated in terms of $\sqrt{p}$, we then have that 
\begin{equation}\label{nablap}
q|\nabla \log \sqrt{p}\, |^2 =  \tan^2\left(\frac{s}{\sqrt{q}}\right) = \frac{1}{p} - 1 \, .
\end{equation}
A brief comment on the statistical interpretation of this expression is in order.  On the ray space, we may define for any observable  $\hat{A}$, the corresponding expectation value function $A(\xi) = \mel{\psi}{\hat{A}}{\psi}$. It is then possible to show (see e.g. \ccite{brodyhugh}) that the uncertainty $\ave{\Delta A^2} = \mel{\psi}{\hat{A}^2}{\psi}-A(\xi)^2$ is related to the gradient of $A(\xi)$ by 
$$
\frac{4}{q}\ave{\Delta A^2} = g^{\mu \nu}(\nabla_\mu A)(\nabla_\nu A) = |\nabla A|^2 \, ,
$$
where $g^{\mu \nu}$ is the inverse to the Fubini-Study metric. Taking $\hat{A}$ to be the projection operator $\hat{\Pi} = \outket{\psi_f}{\psi_f}$, we obtain $\ave{\Pi } = p$ and $
\ave{ \Delta \Pi^2 } = p(1 - p)$. Thus, 
\begin{equation}
 |\nabla  p\, |^2 = \frac{4}{q}\ave{\Delta \Pi ^2} = \frac{4}{q} p (1 - p) \, ,
\end{equation}
which can be seen to follow directly from \rref{nablap}. Therefore, the connection between the transition probability and the Fubini-Study metric is such that the variantion of $p$ with respect to the geodesic distance is, up to a proportionality constant,  the variance in the frequency with which $\ket{\psi_f}$ is obtained given $\ket{\psi}$\ccite{wootters}. 

Let us then proceed to explore a number of consequences that follow from this connection in conjunction with previously obtained results stemming from the generalized Cauchy-Riemann conditions \rref{cauchgen}. Thus far we have seen that from the phase of $\amp{\psi_f}{\psi}$ it is possible to recover the functional dependence of its modulo be means of line integration. It is now easy to show that in the ray space, the modulus of the amplitude can also  be obtained by {\em differentiation} of the phase.  For this we note, as  shown earlier, that from  the generalized Cauchy Riemann conditions and the definition of the quantum metric  it follows that 
$ 
|\nabla \log \sqrt{p} |^2 =|\nabla \eta - \form{A} |^2\, .
$
Using \rref{nablap}  we see therefore that 
$$  
|\nabla \eta - \form{A} |^2 = 1/p -1 \, ,
$$ and hence that the transition probability can also be expressed as
\begin{equation}\label{secondrelprime}
p = \frac{1}{1 + q\, |\nabla \eta - \form{A} |^2} \, . 
\end{equation}
In other words, we see that the amplitude  can  be parameterized entirely in terms of its phase factor according to
\begin{equation}\label{secondrel}
\langle \psif |\psi \rangle = \frac{e^{i \phs} }{\sqrt{1 + q\, |\nabla \eta - \form{A} |^2 }}  \, . 
\end{equation}
 a form that, if the invariant phase gradient is treated as some velocity field as in semi-classical physics, bears a  slight resemblance to the WKB formula $\psi(x) \propto e^{i \eta}/\sqrt{ |\eta'|}$ in one dimension (note however the different powers of $\eta'$ in the radical).  

The resemblance is  sufficiently intriguing to motivate a interpretation of the invariant phase gradient as a sort of velocity field of certain trajectories on the ray space.  This can be done as follows.  From the generalized Cauchy-Riemann condition \rref{cauchgen}, we have that
\begin{equation}
\nabla_\mu \log \sqrt{p} = \Omega_{\mu \nu}V^\nu \, .
\end{equation}
Using \rref{secondrelprime}, we substitute $p = (1 + q V_\lambda V^\lambda)^{-1}$ to obtain
\begin{equation}
-\frac{q}{2} \frac{\nabla_\mu (V_\nu V^\nu)}{1 + q V_\lambda V^\lambda} = \Omega_{\mu \nu} V^\nu
\end{equation}
We now use the fact that
\begin{eqnarray}
\frac{1}{2}\nabla_\mu (  V_\nu V^\nu)  & = &  V^\nu \nabla_\mu V_\nu \nonumber \\
  & = &  V^\nu \nabla_\nu V_\mu +  V^\nu(\nabla_\mu V_\nu - \nabla_\nu V_\mu) \nonumber \\
  & = &  V^\nu \nabla_\nu V_\mu - \frac{2}{q} V^\nu \Omega_{\mu \nu}
\end{eqnarray}
where we have used the fact that $\D \form{V} = \D (\D \eta - \form{A}) = -\D \form{A} = - \frac{2}{q} \form{\Omega}$. Hence we have that
\begin{equation}
 \nabla_\vect{V} V^\mu = \frac{1}{q}\left(1 - q|\vect{V}|^2 \right)\, \Omega^{\mu}{}_\nu V^\nu \, , \, \ \ \ 
\end{equation}
where $\nabla_\vect{V} =V^\nu \nabla_\nu$ is the covariant derivative along the vector field $V^\mu$.  Now note that  because of the anti-symmetry of $\form{\Omega}$, the magnitude of $\vect{V}$ is preserved along its integral lines, i.e., $\nabla_\vect{V} |\vect{V}|^2 = 0$, in consistency with the the fact that the transition probability is constant in the direction of $\vect{V}$. Parameterizing the integral curves of $\vect{V}$ in terms of the geodesic distance along the curve as
$
V^\mu = |\vect{V}| {d\xi^\mu \over ds } \, ,
$
we  obtain the equation 
\begin{equation}
 \frac{ d^2 \xi^\mu}{ds^2}+ \Gamma_{\nu \lambda}^{\mu}{d\xi^\nu \over ds }{d\xi^\lambda \over ds }  = e_{|\vect{V}|}  F^{\mu}{}_\nu {d\xi^\mu \over ds } \, \ \ \ 
\end{equation}
where $e_{|\vect{V}|}$ is a specific constant to each curve given by
$$
e_{|\vect{V}|} = \frac{1 - q|\vect{V}|^2 }{2 |\vect{V}|} \, ,
$$
and $F^{\mu}{}_\nu$ is the field strength associated with the Berry-Simon connection $(\form{F} = \D \form{A})$. From this we see that integral curves of  the invariant phase gradient vector field $V^\mu$  are in correspondence with trajectories on the ray space of charged particles  subject to the magnetic field associated with the Berry-Simon connection. 

Finally, it is worth noting a simplification on the ray space of the scalar integrability condition \rref{laps2} that follows from relation \rref{nablap}, namely
\begin{equation}\label{lapp}
\nabla^2 p = - \frac{4(k + 1)}{q}\left[ p - \frac{1}{k+1} \right] \, .
\end{equation}
Since the ray space is compact, $\int_{\cal R} d\mu \nabla^2 p = 0$, and therefore a volume integration over the entire space of this equation entails that
$$
\langle p \rangle_{\cal R} = \frac{\int_{\cal R} d\mu p}{ \int_{\cal R} d\mu } = \frac{1}{k+1} \, ,
$$
in consistency with the fact that the average of $\outket{\psi}{\psi}$ over the entire ray space should be the completely mixed density matrix of a $k+1$ dimensional Hilbert space.  Equation \rref{lapp} then tells us that the deviation of the transition probability from its average value on the ray space is an eigenfunction of the laplacian operator with eigenvalue $-\frac{4}{q}(k + 1)$. This is easily verified for the Bloch sphere ($k=1$ and choosing $q = 4$), in which case  $p-\frac{1}{2} = \frac{1}{2}\hat{n}\cdot \hat{m}$ is made up of spherical harmonics of order $l=1$.

\section{Geometric Phases}
\label{geoph}

To conclude, we   connect the present results with known results on geometric phases.
As a first application we make a connection with a result of  Samuel and Bhandari  on the Pancharatnam phase. Pancharatnam suggested that an operational definition of what it meant for to quantum states to be ``in" or "out of" phase was naturally provided by the inner product between the two states. The phase $\eta = \arg \amp{\psi_f}{\psi}$ is therefore also called the {\em Pancharatnam phase difference}.  Samuel and Bhandari \cite{sambhand} have shown that this phase has an intrinsic geometric meaning as it can be obtained from the Berry-Simon connection using the geodesic rule, i.e, 
\begin{equation}\label{sambandphase}
\eta = \int\, d{\xi'}^\mu\, A_\mu 
\end{equation}
where the integral is evaluated along the geodesic connecting  the ray $[\ket{\psi_f}]$ with $[\ket{\psi}]$ and where it is assumed that $\ket{\psi_f}$ is an element of the same state section as $\ket{\psi}$. With the aid of the Cauchy-Riemann and the relationship between $p$ and the geodesic distance, it is now seen that for an arbitrary integration path between the two rays \rref{sambandphase} generalizes to
\begin{equation}
\eta = \eta_o + \int\, d\xi^\mu\, A_\mu   + \frac{1}{\sqrt{q}}\int\, d\xi^\mu\, \Omega_\mu{}^\nu\partial_\nu s \,  \tan \frac{ s}{\sqrt{q}}  \, ,
\end{equation} 
where $s$ denotes the geodesic distance from the initial ray to point of integration. If the path of integration is chosen along the geodesic, then from the anti-symmetry of the K\"ahler form the differential $d\xi^\mu \Omega_\mu^\nu \partial_\nu s$ vanishes and equation  \label{sambandphase} is obtained up to the phase relating $\ket{\psi_f}$ with the element of the section $\ket{\psi}$ at $[\, \ket{\psi_f}
 \, ]$.  
 
Next, we turn to geometric phases under time evolution. As is well known, in the course of time evolution the amplitude between the instantaneous state of a system $\ket{\psi;t}$, and the 
initial state $\ket{\psi;0}$ acquires a total phase that can be decomposed into a dynamical phase $-\int dt\, E(t) = -\int dt\, \qave{\psi;t}{\hat{H}(t)}$ and a geometric part. It has been shown by Aharonov and Anandan \cite{anaha87} that when the system undergoes a cyclic evolution so that the  state returns to the initial ray, the geometric contribution to the phase difference acquired is given by $-\oint \form{A}$.  The result generalizes a previous result by Berry\cite{berry}, in which the same geometric phase difference is acquired in the course of adiabatic evolution if the initial state is initially an eigenstate of an adiabatically varying Hamiltonian. 

We now wish to generalize the above results by showing that the phase difference between the exact state $\ket{\psi;t}$ and any arbitrary state $\ket{\psi_f}$ can also be separated into dynamic and geometric contributions and give explicit formulas for the geometric component. The idea then is to consider the phase of an amplitude $\amp{\psi_f}{\psi;t} $
$$
\beta = \arg \amp{\psi_f}{\psi;t} 
$$
where the state $\ket{\psi;t}$ satsifies the evolution equation
$$
i \partial_t \ket{\psi;t} = \hat{H}(t)\ket{\psi;t} \, .
$$
We consider, as an intermediate step,  some arbitrary state section $\ket{\psi(\xi)}$ on the $U(1)$ bundle over the ray space, so that at any given time the time evolved state may be written as
$$
\ket{\psi;t} = e^{i\phi(t)}\ket{\psi(\xi(t))}\, .
$$
Substituting into the Schr\:odinger equation and taking the inner product with $\ket{\psi;t}$, we then find that the phase factor $\phi(t)$ satisfies
\begin{eqnarray*}
\dot{\phi} & =& -\qave{\psi}{\hat{H}(t)} + i \mel{\psi(\xi)}{\partial_t}{\psi(\xi)} \\
 & = & -E(t) - A_\mu \dot{\xi}^\mu \, .
\end{eqnarray*}
Now, letting $\eta = \arg \amp{\psi_f}{\psi}$, we then have
\begin{eqnarray}
\dot{\beta} & = &  \dot{\phi} + \dot{\eta} \\ 
            & = &  -E(t) - A_\mu \dot{\xi}^\mu  + \eta_{,\mu} \dot{\xi}^\mu  \, .
\end{eqnarray}
Thus we recognize the term $  \eta_{,\mu}-A_\mu $ as the invariant phase gradient for the transition amplitude corresponding to the state section in question.  We can then apply the
generalized Cauchy-Riemann conditions to deduce finally that
\begin{equation}
    \dot{\beta} =  -E(t) -\dot{\xi}^\mu \, \Omega_\mu{}^\nu \partial_\nu \log \sqrt{p(\xi)}\, \, .
\end{equation} 
The result shows that in the course of time evolution, the amplitude $\arg \amp{\psi_f}{\psi;t} $ acquires  aside from the dynamical phase, a geometric component
\begin{equation}
\beta_g =  -\int\, d{\xi}^\mu \, \Omega_\mu{}^\nu \partial_\nu \log \sqrt{p(\xi)} \, .
\end{equation}
To see the connection between this result and the cyclic geometric phase we  recall that the curl of  
$
\Omega_\mu{}^\nu \partial_\nu \log \sqrt{p(\xi)}\D \xi^\mu
$
is the  same curl as the curl of $\form{A}$. Thus, when the evolution on the ray space is cyclic, we obtain the usual geometric phase $-\oint \form{A}$. Note also that the present results are can be extended to any complex submanifold of ${\cal R}$ if under  time evolution the state remains on the manifold, for instance by virtue of adiabatic time evolution. 

\section{ Acknowledgements} The author wishes to thank Y. Aharonov, J. Anandan, P. Mazur and E.C.G. Sudarshan for helpful discussions. This work was supported in part by the Office of naval Research under grant no.  N00014-00-1-0383.

\end{document}